\documentclass[conference,11pt]{IEEEtran}
\ifCLASSOPTIONcompsoc
  \usepackage[nocompress]{cite}
\else
  \usepackage{cite}
\fi
\ifCLASSINFOpdf
\else
\fi

\usepackage{hyperref}
\usepackage{url}
\usepackage{graphicx}
\usepackage{booktabs} 
\usepackage{adjustbox}
\usepackage{amsmath}
\usepackage{colortbl} 
\usepackage{xcolor} 
\usepackage[skins]{tcolorbox}

\begin{document}

\title{Speaker Verification using Convolutional Neural Networks}

\author{\IEEEauthorblockN{Hossein Salehghaffari}
\IEEEauthorblockA{Control/Robotics Research Laboratory (CRRL),\\
Department of Electrical and Computer Engineering,\\
NYU Tandon School of Engineering (Polytechnic Institute), NY 11201, USA\\
Email: {h.saleh}@nyu.edu
}

}

\maketitle

\begin{abstract}
In this paper, a novel Convolutional Neural Network architecture has been developed for speaker verification in order to simultaneously capture and discard speaker and non-speaker information, respectively. In training phase, the network is trained to distinguish between different speaker identities for creating the background model. One of the crucial parts is to create the speaker models. Most of the previous approaches create speaker models based on averaging the speaker representations provided by the background model. We overturn this problem by further fine-tuning the trained model using the Siamese framework for generating a discriminative feature space to distinguish between same and different speakers regardless of their identity. This provides a mechanism which simultaneously captures the speaker-related information and create robustness to within-speaker variations. It is demonstrated that the proposed method outperforms the traditional verification methods which create speaker models directly from the background model.

\end{abstract}

\IEEEpeerreviewmaketitle

\section{Introduction}

In speaker verification~(SV), the identity of a query spoken utterance should be confirmed by comparing to the gallery of known speakers. The speaker verification can be categorized to text-dependent and text-independent. In text-independent, no restriction is considered for the utterances. On the other hand, in text-dependent setting, all speakers repeat the same phrase. Due to the variational nature of the former setup, it considers being a more challenging task since the system must be able to clearly distinguish between the speaker and non-speaker characteristics of the uttered phrases. The general procedure of speaker verification consists of three phases: Development, enrollment, and evaluation. For development, a background model must be created for capturing the speaker-related information. In enrollment, the speaker models are created using the background model. Finally, in the evaluation, the query utterances are identified by comparing to existing speaker models created in the enrollment phase.

Recently, with the advent of deep learning in different applications such as speech, image recognition and network pruning~\cite{simonyan2014very,krizhevsky2012imagenet,hinton2012deep,han2015learning},~data-driven approaches using Deep Neural Networks~(DNNs) have also been proposed for effective feature learning for Automatic Speech Recognition~(ASR)~\cite{hinton2012deep} and Speaker Recognition~(SR)~\cite{lei2014novel,variani2014deep}. Also deep architecture has mostly been treated as black boxes, some approaches based on Information Theory~\cite{shannon2001mathematical}, have been presented for multimodal feature extraction and demonstrated promising results~\cite{gurban2009information}.

Some traditional successful model for speaker verification are Gaussian Mixture Model-Universal
Background Model (GMM-UBM) \cite{reynolds2000speaker} and i-vector \cite{dehak2011front}. The main disadvantage of these models is their unsupervised nature since there are not trained objectively for speaker verification setup. Some methods have been proposed to supervise the aforementioned models training such as SVM-based GMM-UBMs \cite{campbell2006support} and PLDA for i-vectors model~\cite{garcia2011analysis}. With the advent of Convolutional Neural Networks~(CNNs) and their promising results for action recognition~\cite{ji20133d}, scene understanding~\cite{tran2015learning}, recently they have been proposed as well for speaker and speech recognition~\cite{variani2014deep,abdel2014convolutional}.

In this work, we propose to use the Siamese neural networks to operate one traditional speech features such as MFCCs\footnote{Mel Frequency Cepstral Coefficients} instead of raw feature for having a higher-level representation for speaker-related characteristics. Moreover, we show the advantage of utilizing an effective pair selection method for verification purposes.

\section{Related Works}

Convolutional Neural Networks \cite{lecun1998gradient} have recently been used for speech recognition~\cite{sainath2013deep}. Deep models have effectively been proposed an utilized for text-independent setup in some research efforts~\cite{lei2014novel,richardson2015deep}. Locally Connected Networks (LCNs) have been utilized for SV as well~\cite{chen2015locally}.~Although in \cite{chen2015locally}, the setup is text-dependent. In some other works such as \cite{heigold2016end,zhang2017end}, the deep networks have been employed for feature extractors to create speaker models for further evaluations. We investigate the CNNs specifically trained end-to-end for verification purposes and furthermore employ them as feature extractors to distinguish between the speaker and non-speaker information.

\section{Speaker Verification Procedure and Protocol}

\begin{figure}[!tbh]
\begin{center}
   \includegraphics[width=0.8\linewidth]{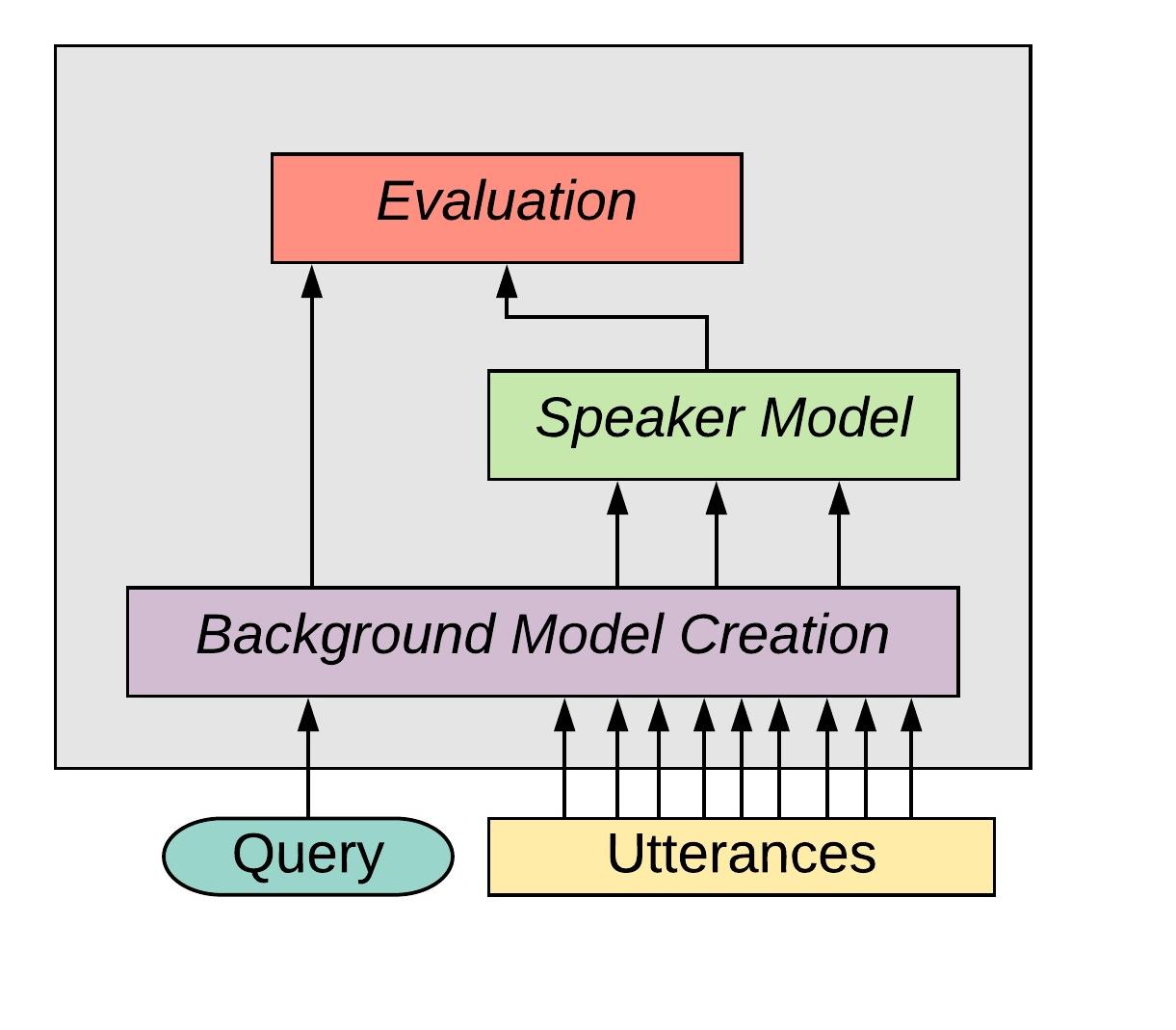}
\end{center}
   \caption{General view of the speaker verification.}
\label{fig:SV}
\end{figure}

The speaker verification protocol can be categorized into three phases: development, enrollment, and evaluation. The general view of the speaker verification protocol is depicted in Fig~\ref{fig:SV}. We explain these phases in this section with a special emphasis on how they can be adapted to deep learning. Different research efforts proposed variety of methods for implementing and adapting this protocol such i-vector~\cite{dehak2011front,kenny2007joint}, d-vector system~\cite{variani2014deep}.\\

\noindent \textbf{Development } In the development stage, speaker utterances are utilized for creating a background model for speaker representation. Different elements such as representation level~(frame or utterance-level), the model type such as deep networks or Bayesian models and training objective~(loss function) forms the speaker representation type. The main motivation behind employing DNN is to use their architecture as a powerful speaker feature extractor.\\

\noindent \textbf{Enrollment } In this stage, a distinct model should be created for each speaker identity. The speaker utterances will be utilized for speaker model generation. In the case of DNNs, is this phase, speaker utterances will be the input to the model created by previous phase and the outputs will be integrated with some method to create the unique speaker model. The speaker representation provided by averaging the outputs of the DNN~(called d-vectors) is a common choice~\cite{chen2015locally,variani2014deep}. \\

\noindent \textbf{Evaluation } During the evaluation phase, test utterances will be fed to the model for speaker representation extraction. The query test sample will be compared to all speaker models using a score function and the one with the highest score is the predicted speaker. Considering the one-vs-all setup, this stage is equivalent a binary classification problem in which the traditional Equal Error Rate~(EER) is used for model evaluation. The false reject rate and the false accept rate are determined by predefined threshold and when two errors become equal, the operating point is EER. Usually, as for the scoring function, the simple \textit{cosine similarity score} will be employed. The score is measured by the similarity between the representation of the test utterance and the targeted speaker model.

\section{Dataset}

We used the public VoxCeleb dataset for our experiments~\cite{Nagrani17}. It contains around 140k utterances for 1211 speakers and around 6k utterances for 40 speaker identities used for testing. The dataset is balanced based on the gender and has been spanned through different ethnicities and accents. The audios are extracted from uploaded videos to YouTube and have been captured in a wide variety of challenging multi-speaker settings including background chatter, overlapping speech, channel noise and different qualities of recording. The general statistics of the dataset are available in Table~\ref{table:VoxCelebdataset}.

\begin{table}[h]
\caption[Table caption text]{Statistics of the VoxCeleb dataset.}
\label{table:VoxCelebdataset}
\begin{center}
\addtolength{\tabcolsep}{-3pt}
\begin{tabular}{ccc}
\toprule 
\# of & Development  & Test \\
\hline
\midrule
\rowcolor{black!0} Speakers & 1,211 & 40 \\ 
\rowcolor{black!0} Videos  & 21,819 & 677 \\
\rowcolor{black!0} Utterances & 139,124 & 6,255 \\

\bottomrule
\end{tabular}
\end{center}

\end{table}

\section{Architecture}

The aim is to utilize CNNs as powerful feature extractors. The input pipeline and the specific architecture are explained in this section.

\subsection{Input pipeline}

The audio raw features are extracted and re-sampled to 16kHz. For spectrogram generation, 25ms hamming windows with a step size of 10ms are used for 512-point FFT spectrum. We used 1-second of audio stream and it yields to a spectrogram of size $256 \times 100$.~No mean or variance normalization has been used. No voice activity detection has been performed as well.~Out of the generated spectrum, 40 log-energy of filter banks per hamming window alongside their first and second order derivatives are generated to form $3 \times 40 \times 100$ input feature map.~For feature extraction, the SpeechPy library~\cite{torfi2018speechpy} has been utilized.

\subsection{Architecture Design}

We use an architecture similar to VGG-M~\cite{chatfield2014return}, widely used for image classification and speech-related applications~\cite{Chung16a}. The details are available in Table~\ref{table:architecture}. We modified the architecture with some considerations: (1)~It should be adapted to our input pipeline, (2)~we do not use any pooling layer and (3)~the size has been shrunk for having a smaller architecture for faster training and empirically we found it to be less prone to overfitting. The main reason for not pooling in time is to keep the temporal information although it is claimed that it may increase the robustness to temporal variations. In practice we found it degrading the performance to perform pooling in the time dimension. Our observations have been further investigated an verified by \cite{abdel2014convolutional} as well.

\begin{table}[h]
\caption[Table caption text]{The architecture used for verification purpose.}
\label{table:architecture}
\begin{center}
\addtolength{\tabcolsep}{-3pt}
\begin{tabular}{ccccc}
\toprule 
Layer & Kernel & \# Filters  & Stride & Output size \\
\hline
\midrule
\rowcolor{black!0} Conv-1 & $7 \times 7$ & 32 & $2 \times 2$ & $32 \times 17 \times 47$ \\ 
\rowcolor{black!0} Conv-2  & $5 \times 5$ & 64 & $1 \times 1$ & $64 \times 13 \times 43$  \\
\rowcolor{black!0} Conv-3 & $3 \times 3$ & 128 & $1 \times 1$ & $128 \times 11 \times 41$  \\
\rowcolor{black!0} Conv-4 & $3 \times 3$ & 256 & $1 \times 1$ & $256 \times 9 \times 39$  \\
\rowcolor{black!0} Conv-5 & $3 \times 3$ & 256 & $1 \times 1$ & $256 \times 7 \times 37$  \\
\rowcolor{black!0} fc-1 & - & 1024 & - & -  \\
\rowcolor{black!0} fc-2 & - & 256 & - & -  \\
\rowcolor{black!0} fc-3 & - & 1251 & - & -  \\

\bottomrule
\end{tabular}
\end{center}

\end{table}

\subsection{The Verification Setup}

A usual method is to train a Softmax loss function for classification and use the features extracted by the fully-connected layer prior to Softmax.~However, a reasonable criticism about this method is that the Softmax loss criterion tries to identify speakers and not verify the available identity in a one-vs-all setup inconsistent with the speaker verification protocol. Instead, we utilize a Siamese neural network as proposed in \cite{chopra2005learning} and implemented in many research efforts~\cite{sun2018deep,varior2016gated,koch2015siamese}.

\begin{figure}[!tbh]
\begin{center}
   \includegraphics[width=0.8\linewidth]{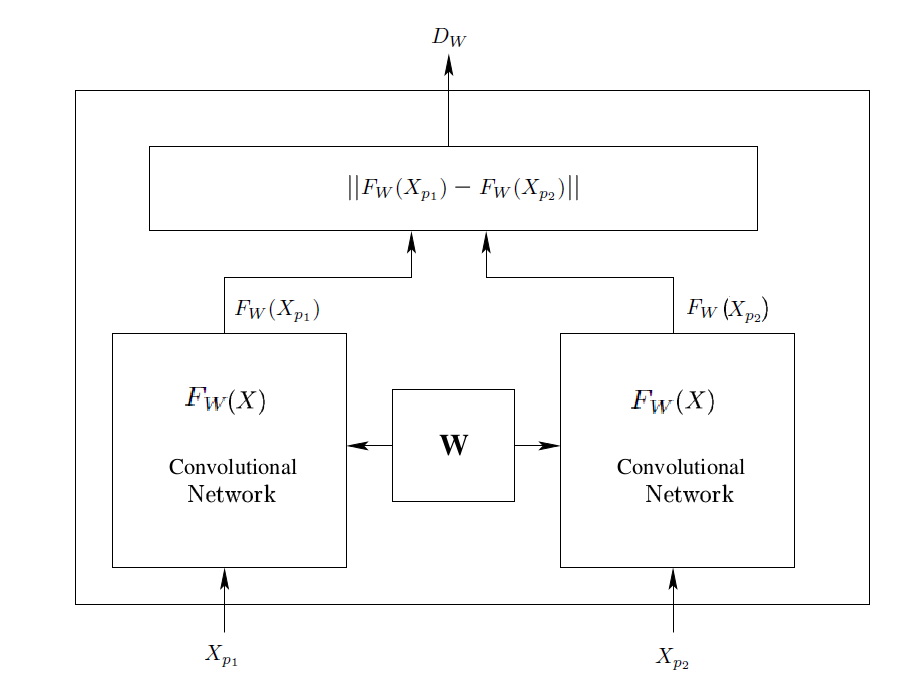}
\end{center}
   \caption{Siamese Model Framework}
\label{fig:Model}
\end{figure}

The Siamese architecture consists of two identical CNNs. The main goal is to create a common feature subspace for discrimination between match and non-match pairs based on a distance metric. The model is demonstrated in Fig.~\ref{fig:Model}. The general idea is that when two pairs belong to the same identity, their distance in the common feature subspace should be as close as possible and as far as possible, otherwise. Assume $X_{p_{1}}$ and $X_{p_{2}}$ are the input pairs of the system and the distance between a pair of input in the output subspace defined as $D_W(X_{p_{1}},X_{p_{2}})$ (i.e., the $\ell_2-norm$ between two vectors), then the distance is computed as follows:


\begin{equation}\label{eq1}
D_W (X_{p_{1}},X_{p_{1}}) = {||F_W (X_{p_{1}})-F_W (X_{p_{2}})||}_{2} .
\end{equation}

The system will be trained using contrastive cost function. The goal of contrastive cost $L_W(X,Y)$ is to minimize the loss in both scenarios of encountering match and non-match pairs, with the following definition:
\begin{align}\label{eq2}
L_W(X,Y) = {{1}\over{N}}  \sum_{i=1}^{N} L_W(Y_i,(X_{p_{1}},X_{p_{2}})_i),
\end{align}

\noindent where N is the number of training samples, $i$ is the indication of each sample and $L_W(Y_i,(X_{p_{1}},X_{p_{2}})_i)$ is defined as follows:

\begin{equation} \label{eq3}
\begin{split}
L_W&(Y_i,(X_{p_{1}},X_{p_{2}})_i) = Y*L_{gen}(D_W(X_{p_{1}},X_{p_{2}})_i)\\ &+ (1-Y)*L_{imp}(D_W(X_{p_{1}},X_{p_{2}})_i)+\lambda{||W||}_{2}
\end{split}
\end{equation}

\noindent in which the last term is the regularization. $L_{gen}$ and $L_{imp}$ will be defined as the functions of $D_W(X_{p_{1}},X_{p_{2}})$ by the following equations:

\begin{equation}\label{eq4}
  \begin{cases}
    L_{gen}(D_W(X_{p_{1}},X_{p_{2}}))={{1}\over{2}}{D_W(X_{p_{1}},X_{p_{2}})}^2\\
    L_{imp}(D_W(X_{p_{1}},X_{p_{2}}))={{1}\over{2}}max\{0,{(M-D_W(X_{p_{1}},X_{p_{2}}))}\}^2
  \end{cases}
\end{equation}

\noindent where $M$ is a predefined margin.~Contrastive cost is implemented as a mapping criterion which supposed to put match pairs to nearby and non-match pairs to distant places in the output manifold. 


%

\section{Implementation}

We used TensorFLow~\cite{tensorflow2015-whitepaper} as the deep learning framework and our model has been trained on a NVIDIA Pascal GPU. No data augmentation has been used for the development phase. Batch normalization has been employed for having the robustness to s internal covariate shifts and being less affected by initialization~\cite{ioffe2015batch}. For verification, after training the network as a classifier~(initial learning rate = 0.001), we fine-tune the network by training the Siamese architecture~(with an initial learning rate of 0.00001) for 20 epochs. Unlike the procedure used by~\cite{Nagrani17}, we do not freeze the weights of any layer for fine-tuning.

\section{Experiments}

In this section, we describe the experiments performed for speaker verification and we compare our proposed method to some of the existing methods.

\subsection{Experimental Setup}

For speaker verification, we followed the protocol provided by~\cite{Nagrani17} and the identities their names start with \textbf{'E'} are used for testing. The subjects are not used for purpose of training and testing though. We only use the subjects to create match and non-match pairs for verification purposes. As it has been mentioned, the performance metric used in our evaluation phase is the EER that is commonly used for verification systems.

\subsection{Methods}

DIfferent methods have been implemented and used for comparison.

\noindent \textbf{GMM-UBM} For GMM-UBM method~\cite{reynolds2000speaker}, MFCCs with 40 coefficients with cepstral mean and variance normalisation are used. For training the Universal Background Model~(UBM) made of 512 mixture components, 20 iterations of the training data have been used.

\noindent \textbf{I-vectors} The I-vectors system has been widely known as one of the state-of-the-art representations which operate as frame-level as proposed in~\cite{dehak2011front}. Probabilistic linear discriminant analysis~(PLDA) has also been used on top of I-vector for dimensionality reduction~\cite{kenny2010bayesian}.

\subsection{Results}

We used different dimensions for our embedding level. Moreover, we used I-vector with and without PLDA to showcase the effect of probabilistic dimensionality reduction.

\begin{table}[h]
\caption[Table caption text]{The architecture used for verification purpose.}
\label{table:architecture}
\begin{center}
\addtolength{\tabcolsep}{-3pt}
\begin{tabular}{cc}
\toprule 
Model & EER\\
\hline
\midrule
\rowcolor{black!0} GMM-UBM & 17.1 \\ 
\rowcolor{black!0} I-vectors   & 12.8 \\
\rowcolor{black!0} I-vectors + PLDA~\cite{dehak2011front}  & 11.5 \\
\rowcolor{black!0} CNN-2048  & 11.3 \\
\rowcolor{black!0} CNN-256 + Pair Selection  & 10.5 \\

\bottomrule
\end{tabular}
\end{center}

\end{table}

\section{Conclusion}

We proposed an end-to-end architecture alongside with adapting active learning procedure for pair selection for speaker verification application. It is observed the effective online pair selection method in addition to training the system in an end-to-end fashion, can outperform the traditional method that uses the background models for speaker representation. A proposed CNN architecture has also been trained as a feature extractor on top of the traditional speech features rather than the raw audio for directly capturing the inter-speaker and intra-speaker variations.


%
%

\bibliographystyle{ieeetr}
\bibliography{ref}

\end{document}